\documentclass{article}
\usepackage{amsmath, amssymb,graphicx}
\usepackage{hyperref}
\newcommand{\mathsym}[1]{{}}

\newcommand{\address}{{\small \v{Z}ilinsk\'a univerzita, Univerzitn\'a 1, 010\:26 \v{Z}ilina, Slovakia}}

\begin{document}
\title{Higgs potential and fundamental physics}
\author{Ivan Melo \\ \address}

\maketitle

\begin{abstract}
Physics associated with the Higgs field potential is rich and interesting and deserves 
a concise summary for a broader audience to appreciate
the beauty and the challenges of this subject. We discuss the role of the Higgs potential in particle physics, in particular in the spontaneous symmetry breaking
and in the mass generation using an example of a simple reflection symmetry, then continue with the temperature and quantum corrections to the potential which lead us to the naturalness problem and the vacuum stability.

\end{abstract}

\section{Introduction}
\label{intro}
Since the discovery of the Higgs boson by the ATLAS and CMS collaborations at the LHC collider at CERN in 2012 \cite{Discovery}, we have to
take the existence of the scalar Higgs field very seriously. Its unique role in the Standard model of elementary particles (SM) draws the attention of many people, including those outside of particle physics.
Naturally, there is an ongoing effort, including this journal \cite{EJP}, to explain the subject at a level suitable for nonspecialists. These papers typically concentrate on the key concept of electroweak symmetry and its spontaneous breaking (Higgs mechanism) and aim at audiences beginning with high school or undergraduate students. Here we focus on the Higgs potential and its shape in order to summarize and discuss the associated many faceted (and fascinating) physics.
With this goal in mind, the symmetry of the Higgs potential is reduced to a discrete reflection for simplicity.
The paper is aimed at university physics teachers who do not work within elementary particle physics. Other audiences could benefit from this review to the degree they know theoretical physics, in particular
quantum mechanics and at least basics of quantum field theory and particle physics.

We first discuss the Landau theory of phase transitions, then address the basics of the Higgs potential, the importance of the Higgs field for masses of elementary particles, the cosmological constant problem, the electroweak phase transition, the naturalness problem and finally the vacuum stability.

\section{Landau theory of phase transitions}
\label{Landau}
Motivation for the Higgs potential comes from the Landau theory of phase transitions. We will consider a ferromagnetic phase transition as an example. 
In a toy model the ferromagnet consists of atoms arranged in a plane, each of which has a magnetic moment constrained to point either up or down \cite{Mee}.
Ferromagnetic materials lose their magnetic properties at temperatures higher than the critical temperature $T_c$. 
In this high temperature phase the system is disordered as magnetic moments point at random up or down due to the thermal motion of atoms. The net magnetization $M$ is zero since on average the number of magnetic moments pointing up equals the number pointing down. On the other hand, when cooled below $T_c$, the magnet develops nonzero magnetization as atoms prefer to align magnetic moments with their neighbours and, as a result, all moments point either up or down (system becomes ordered and $M$ is the order parameter). Either of the two directions is allowed with the same probability since magnetic interactions between neighbouring atoms are symmetric with respect to up $\leftrightarrow$ down reflection.

Following Landau we write the energy of this system in the vicinity of $T_c$ as an expansion in even powers of the (small) order parameter $M$ with unknown coefficients $\alpha, \beta$,
\begin{equation}
\label{freeenergy}
E(M) = \alpha M^2 + \beta M^4.
\end{equation}
Higher order terms with even powers of $M$ could be added in principle but not necessarily, the odd power terms are ruled out due to the symmetry requirements (they violate up $\leftrightarrow$ down symmetry). The coefficient $\alpha$ depends on temperature - it is positive for $T > T_c$, zero for $T = T_c$  and negative for $T < T_c$; $\beta > 0$ in both phases. 
The sign of $\alpha$ leads to the crucial difference between the two phases: the energy for $T > T_c$ has a single minimum at $M = 0$, while for $T < T_c$ there are two minima at $M = \pm t$ where $t = \sqrt{\frac{-\alpha}{2 \beta}}$  (Fig. 1). Above $T_c$ the system is in the minimum with zero magnetization, below $T_c$ it spontaneously falls into one of the two minima with the nonzero order parameter $M$. 
\begin{figure}[htb]
\begin{center}
\includegraphics[height=1.65in,width=3.0in]{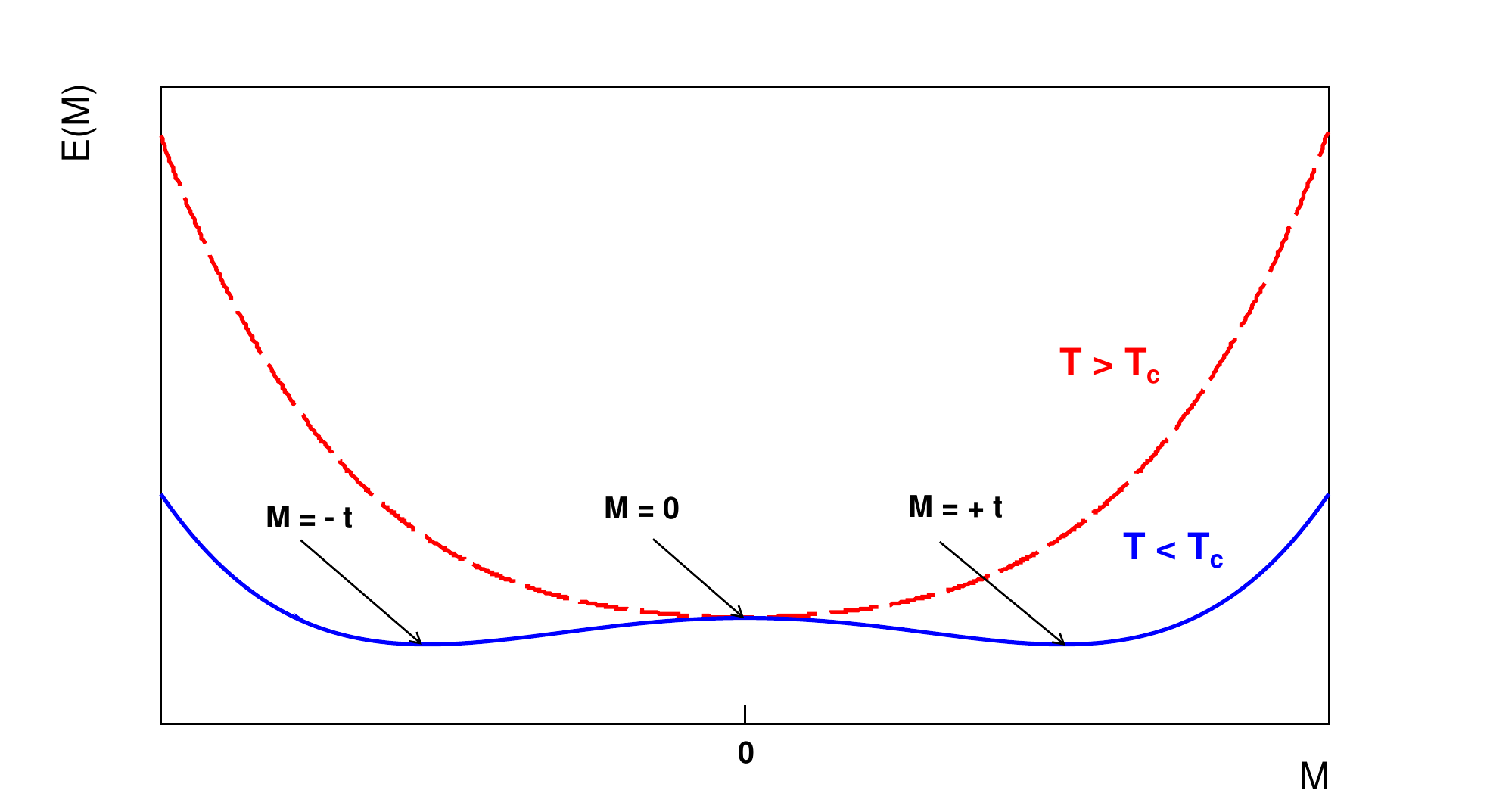} 
\caption{Energy of a ferromagnet as a function of the magnetization $M$. For $T > T_{c}$ (dashed red line) the minimum is at $M = 0$. For $T < T_{c}$ (solid blue line) the symmetry is spontaneously broken and the magnet randomly chooses one of the two minima $M = \pm t$.}
\label{ferromagnet_fig}
\end{center}
\end{figure}

There is also an important difference in symmetry between the two phases. The high temperature minimum energy state has $M = 0$ which is symmetric with respect to 
$M \leftrightarrow -M$ (i.e., after reflection we end up in the same $M = 0$ state) but in the low temperature state the reflection symmetry is broken: the minimum energy state $M = t$ (all magnetic moments pointing up) is reflected to the second minimum with $M = -t$ (all moments pointing down). 
On the other hand the energy in Eq.\ref{freeenergy} is symmetric with respect to reflection 
$M \leftrightarrow -M$ for both phases ($E(M) = E(-M)$).
This kind of symmetry breaking, when the energy (given by underlying interactions) is symmetric but the state of the minimum energy is not, is called spontaneous symmetry breaking. The up $\leftrightarrow$ down symmetry of interactions between individual magnetic moments (manifested in Eq. \ref{freeenergy}) is hidden when we observe all of them pointing up (or down).
The loss of symmetry is accompanied by a gain in order in the system as measured by the order parameter $M$.

Landau theory proved to be very successful for a broad class of phase transitions with one of the highlights being the Landau-Ginzburg theory of superconductivity.

\section{Classical Higgs potential}
\label{classical}
Landau's ideas were introduced into particle physics in the 1960s. 
Peter Higgs showed an example of how a symmetry like the symmetry of electromagnetic interactions\footnote{$U(1)$ gauge symmetry. The gauge symmetries are key concept in quantum field theory.} 
could be spontaneously broken 
 if one assumes the existence of a special field (now known as the Higgs field) which uniformly fills the entire Universe and is {\it nonzero} on average even in completely empty space - in the vacuum. This model was just an illustration since the electromagnetic symmetry is in fact not broken, but it proved to be a breakthrough idea. At the time there was a conflict between the symmetry needed to describe weak interactions and the masses of W and Z bosons (the mediators of the weak force) which seemed to violate this symmetry. Weinberg and Salam used the Higgs's ideas to show that 
the weak symmetry\footnote{The weak symmetry in their treatment is deeply interconnected with the electromagnetic symmetry in the form of $SU(2)_L \times U(1)_Y$ gauge symmetry, hence we will use the term electroweak symmetry and electroweak phase transition for the remainder of the text.} is spontaneously broken - weak interactions respect the symmetry but the state of the minimum energy, the vacuum, does not. W and Z boson masses just appear to violate the electroweak symmetry (like the ferromagnet in the low temperature state appears to violate the reflection symmetry) and the conflict is resolved.

\begin{figure}[htb]
\begin{center}
\includegraphics[height=1.65in,width=2.0in]{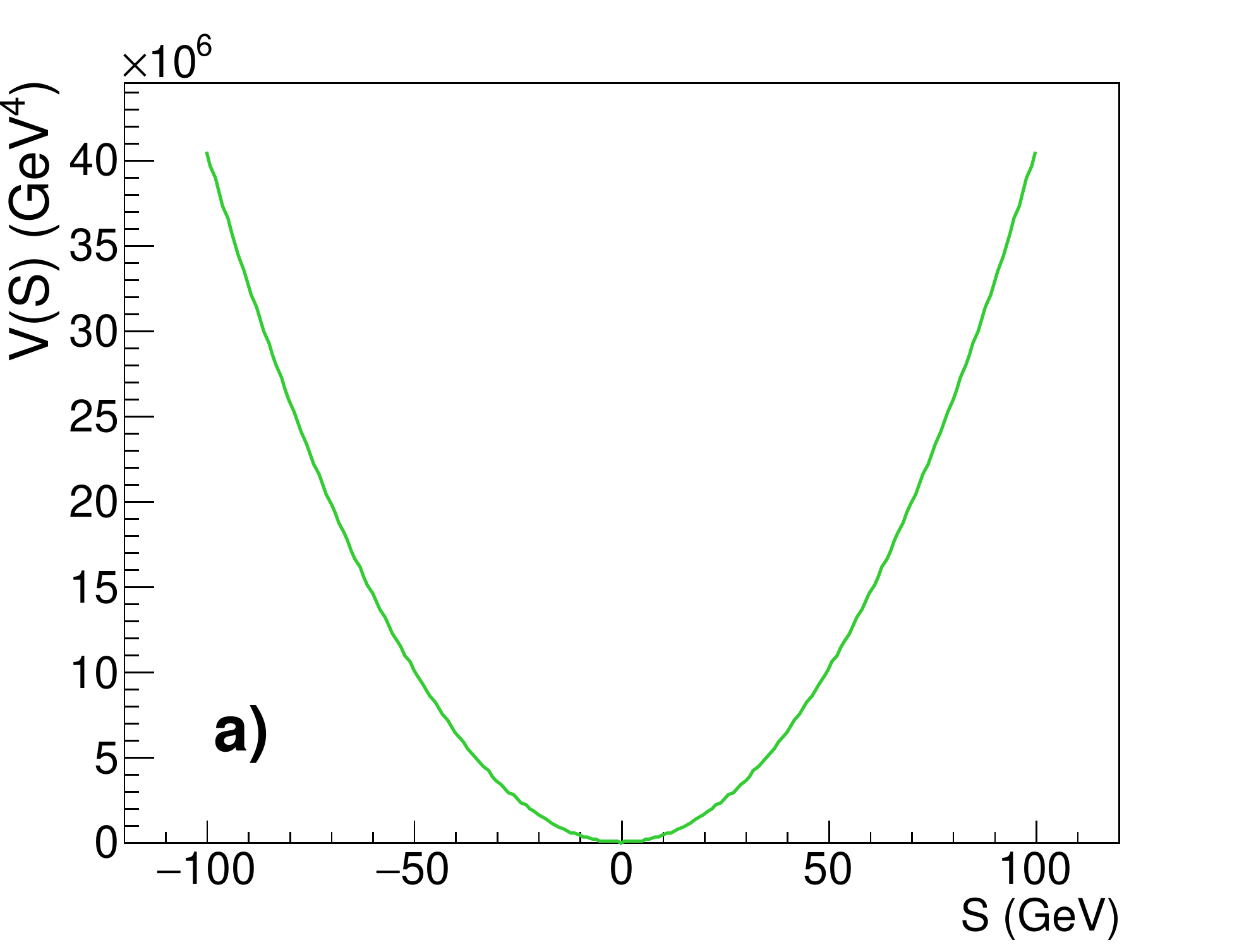}   \includegraphics[height=1.7in,width=2.7in]{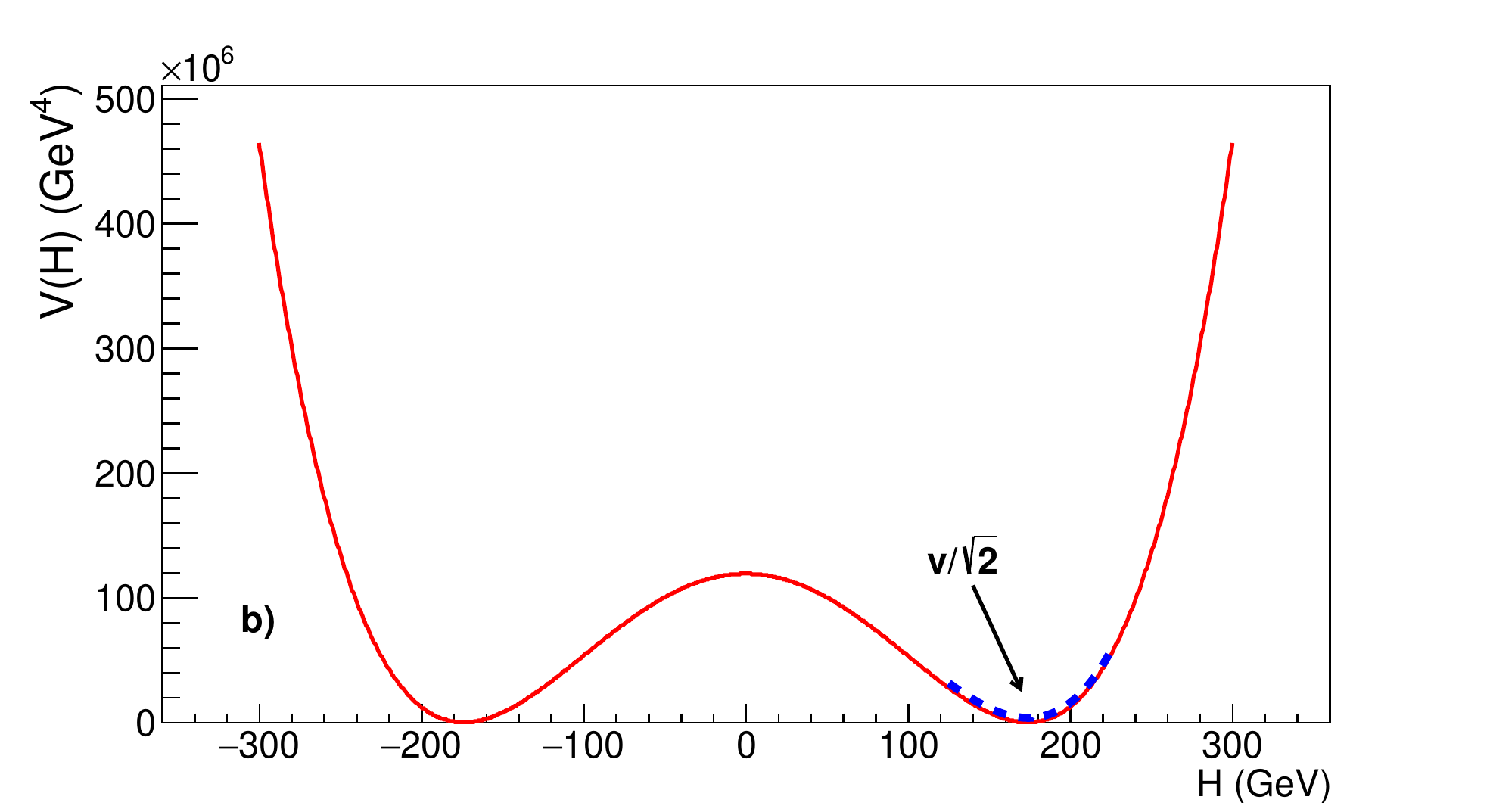}
\caption{a) Potential $V(S)$ as a function of the average value of the field~$S$. b) Higgs potential as a function of the Higgs field average value $H$ for $\mu^2 = - (90\;{\rm GeV})^2$, $\lambda = 0.13$, $c_0 = \mu^4/(4 \lambda)$ (solid red line). Experiments probe the minimum of the potential and its curvature at the minimum (dashed blue line).}
\label{Higgs_potential_fig}
\end{center}
\end{figure}

Let us see why the Higgs field is so important.
According to quantum field theory (QFT) the vacuum is not simply 'nothing' but a complex dynamic state full of fields, one for each elementary particle. These fields change constantly due to quantum fluctuations. The average values of the fluctuating fields in the vacuum are typically zero - the value at which the corresponding potential energy densities, given by Eq. \ref{S_potential}, are at their minima\footnote{To keep the notation and discussion simple, we represent all fields as scalars. The reader should note, however, that only the Higgs field is a true scalar (spin $0$) and this is precisely why its average value in the vacuum can be nonzero. For quark and lepton fields (spin $1/2$) this is not possible and for photon, W, Z fields (spin $1$) we do not observe nonzero average values.}:
\begin{eqnarray}
\label{S_potential}
V(S) & = & \frac{1}{2} m_S^2 \:S^2,
\end{eqnarray} 
where $S$ is the generic massive field and $m_S$ its mass. 
The potential energy density is parabolic with its minimum at the zero average value of the field (Fig.\ref{Higgs_potential_fig}a). The potential energy density and the field are expressed in Natural Units\footnote{Natural Units use the unit of energy GeV, the unit of action (the reduced Planck constant) $\hbar = \frac{h}{2\pi} = 1$ and the speed of light in the vacuum $c = 1$ instead of kg, m and s. Energy, mass, momentum and scalar fields are then expressed in terms of GeV, length and time in terms of GeV$^{-1}$ and potential energy density in terms of GeV$^4$.}.

While quantum fluctuations are irregular, we can make the field fluctuate about its average value in a regular manner like a harmonic wave by adding at least a minimum amount of energy (equal to $m_S$) to it. This harmonic fluctuation is what we know as particle $S$ of mass $m_S$, the quantum of the field $S$. The basic object is the field and the particle is its manifestation. 
We would like to stress that for scalar fields the mass of the particle squared is proportional to the coefficient of the field squared term in the potential energy density (Eq.~\ref{S_potential}). The positive sign of the term is also important as we will see below.

The Higgs field with its nonzero average value in the vacuum, the same across the Universe, is unique among quantum fields and this leads to far reaching consequences to be discussed below. The Higgs potential energy density\footnote{The potential energy density will be called “potential” for the remainder of the text.} $V(H)$ is no longer a simple parabola. Inspired by Landau theory, for a simplified case of a real Higgs field $H$, it can be written as
\begin{eqnarray}
\label{Higgs_potential}
V(H) & = & \mu^2 H^2 +  \lambda H^4 + c_0
\end{eqnarray} 
where $\mu^2$ is the {\it negative} mass squared parameter, $\lambda > 0$ is the Higgs field self-coupling  and $c_0$ is an arbitrary normalization constant independent of $H$ which has no physical consequences as long as we remain within the SM where only differences in energy are important. 

The potential looks like a 'Mexican hat' with a local maximum at $H = 0$ and two minima
at $H = \pm \:v /\sqrt{2}$ where $v^2 = -\mu^2/\lambda$, see Fig.\ref{Higgs_potential_fig}b.
The analogy with the ferromagnet in the low temperature phase $T < T_c$ (Fig. \ref{ferromagnet_fig}) is obvious. The Higgs field (just as the magnetization $M$ for the ferromagnet) plays the role of the order parameter.
The Higgs potential  is symmetric with respect to $H \leftrightarrow -H$ reflection, however, this symmetry is spontaneously broken by the lowest energy state ($H = v /\sqrt{2}$ is reflected to $H = -v /\sqrt{2}$). According to this picture the Universe might have gone through a phase transition very early after the Big Bang from a high temperature phase with the symmetric vacuum (zero average Higgs field) to a low temperature phase\footnote{The temperature of the current Universe is effectively zero.} when it fell into one of the two minima, e.g. $H_{vac} = v /\sqrt{2} > 0$, which became our current vacuum. This phase transition, known as the electroweak phase transition, will be discussed
in Section~\ref{EPT}. 

The Higgs particle (Higgs boson or Higgs in short) is the quantum of the Higgs field. Its mass is defined by the Higgs potential but while
the mass of the $S$ particle is clearly visible in Eq.\ref{S_potential} as the positive coefficient of the $S^2$ term, 
the same is not true for the Higgs since the coefficient $\mu^2$ of the $H^2$ term has the wrong sign, it is negative.  
In order to expose the mass of the corresponding Higgs, 
we have to separate the field $H$ into two parts: a part $h$ corresponding to the Higgs particle, 
which fluctuates about $H_{vac} = v /\sqrt{2}$, and the constant part $v /\sqrt{2}$ itself,
\begin{eqnarray}
H & = & (h + v)/\sqrt{2}.
\end{eqnarray} 
Plugging the separated field $H$ into Eq.\ref{Higgs_potential}, we get for $c_0 = \mu^4/(4 \lambda)$
\begin{eqnarray}
\label{Higgs_mass1} 
V(h) & = & \frac{1}{4} \lambda h^4 + \lambda v h^3 + \lambda v^2 h^2
\end{eqnarray} 
We will focus on the last term, $\lambda v^2 h^2$, which has the form of a mass term, $M_h^2\: h^2/2$, with the correct sign. It corresponds to the Higgs boson with the mass
\begin{eqnarray}
\label{Higgs_mass2} 
M_h^2 & = & 2 \lambda v^2 \;=\; -2\mu^2
\end{eqnarray}
The most important fact that the Higgs field must be nonzero in the vacuum (including the value of $v$) has been known for decades, but we had to wait for its manifestation in the form of the particle until 2012 when the Higgs boson was discovered at the LHC collider with the mass $M_h = 125$ GeV.

The parameters $v, \lambda$ of the Higgs potential (or equivalently $\mu, \lambda$) are determined through the measurement of the Fermi constant\footnote{The Fermi constant $G_F$, the strength of the weak interaction in the Fermi theory, is determined from the measurement of the muon lifetime.} $G_F$ and the Higgs mass, yielding
\begin{eqnarray}
\label{lambda_v}  
v & = & \sqrt{\frac{1}{\sqrt{2}G_F}} \;\doteq\; 246\;{\rm GeV}, \;\;\;\;
\lambda = \frac{M_h^2}{2v^2} \;\doteq\; 0.13  \;\;\;\; \mu^2 \doteq - (90\;{\rm GeV)^2} 
\end{eqnarray}
In fact, what we observe experimentally is just the location of the minimum and the curvature\footnote{The curvature of the Higgs potential at the minimum $H = v/\sqrt{2}$ is defined by the Higgs mass squared as one can see from $\frac{\partial^2 V(H)}{\partial H^2}|_{H = v/\sqrt{2}} = 4 \lambda v^2 \equiv 2 M_h^2$.} of the potential at the minimum, 
indicated as the dashed blue line in Fig.\ref{Higgs_potential_fig}b. 

We shall call the potential $V(H)$ of Eq.\ref{Higgs_potential} 
the {\it classical} potential 
(the solid red line in Fig.\ref{Higgs_potential_fig}b). The classical potential receives important quantum and temperature corrections to be discussed later.

\section{Higgs field and masses of elementary particles}
The classical Higgs potential is closely associated with the problem of the $W$ and $Z$ boson masses which appear to break the electroweak symmetry. 
The spontaneous symmetry breaking of the Higgs sector which we  described above for a toy symmetry $H \leftrightarrow -H$, 
applies also to the electroweak symmetry if the Higgs field interacts with the weak fields. In this way the weak interactions respect the electroweak symmetry
 and at the same time the lowest energy state, the vacuum, breaks it and generates $W$ and $Z$ masses. 
The framework which describes this process is the famous Higgs mechanism of electroweak symmetry breaking \cite{HiggsMechanism}. 
The mechanism itself is beyond the scope of this review, however, an important part can be shown without going into details of the electroweak symmetry. 
This part also applies to massive quarks and leptons, not just $W$ and $Z$ bosons.

Let us consider the field $S$ of the previous section as a generic example of particle mass generation and assume it interacts with the Higgs field $H$. 
Interactions in QFT are described by the products of the fields and the strength of the interaction is given by the coupling constant.
In our case the potential $U$ of the two interacting fields is given by 
\begin{eqnarray}
\label{Higgs_potential_int}
U & = &  y^2 S^2 H^2 + V(H)
\end{eqnarray} 
where $y$ is the coupling constant. When we now express the field $H$ as before, $H = (h + v)/\sqrt{2}$, we get
\begin{eqnarray}
\label{Higgs_mass1_int} 
U  & = & \frac{1}{2}y^2 v^2 S^2 + y^2 v S^2 h + \frac{1}{2} y^2 S^2 h^2 + V(h)
\end{eqnarray} 
The term $\frac{1}{2} y^2 v^2 S^2$ has the form of a mass term, $V(S) = \frac{1}{2} m_S^2 S^2$ (see Eq.\ref{S_potential}), which means that the particle $S$, the quantum of the field $S$, acquired mass
\begin{eqnarray}
m_S &=& y v.
\end{eqnarray} 
Different particles have different couplings $y_i$ with the Higgs field and hence they acquire different masses $m_i$.
The couplings $y_i$ are not predicted by the SM. For $W$ and $Z$ bosons they are given in terms of the coupling strengths of weak interactions (as described in the Higgs mechanism), for quarks and leptons they were introduced on an {\it ad hoc} basis which represents a deep open problem of the SM: we have too many unexplained parameters.

We emphasize two ingredients crucial in this mass generating mechanism: the nonzero vacuum value of the Higgs field, $H_{vac} = v/\sqrt{2}$ and 
the interaction of the field $S$ with the Higgs field $H$.

\section{Cosmological constant problem}
The Higgs potential $V(H)$ can be interpreted as the Higgs contribution to the vacuum energy density (energy density of empty space). 
We recall that it is defined up to a constant. With a choice of $c_0 = \mu^4/(4 \lambda)$ in Eq.\ref{Higgs_potential} and for the average value $H_{vac} = v/\sqrt{2}$ we get 
for the current vacuum energy density due to the Higgs field $V_{vac} = V(v/\sqrt{2}) = 0$. A different choice of $c_0$ gives nonzero $V_{vac}$.

This arbitrariness is fine in the SM, however, if we include gravity into consideration, the constant $c_0$ can no longer be arbitrary. Gravity 'feels' the vacuum energy (also known as the cosmological constant) which means that $V_{vac}$ could affect the evolution of the Universe \cite{Rugh}.
While it is not clear what the correct value of $V_{vac}$ is, the expectation is that it is large \cite{Rugh,Vulpen}, 
of the order of the Higgs potential at $H = 0$ in Fig.\ref{Higgs_potential_fig}b,
\begin{eqnarray} 
V_{vac} & \sim & \frac{1}{4}\lambda v^4 = 1.2 \times 10^{8}\: {\rm GeV}^{4} \doteq 10^{44}\: {\rm eV}^4.
\end{eqnarray} 
The problem is that the vacuum energy density observed in cosmology is  \cite{Cosmological_constant}
\begin{eqnarray}
V_{cosm} &=& (0.003)^4 {\rm eV}^4 \sim 10^{-10} {\rm eV}^4,
\end{eqnarray}
smaller by a stunning factor of $\sim 10^{54}$. 
This huge difference between theory and observation is a mystery: an extreme fine-tuning is required between the Higgs $V_{vac}$ and other sources of vacuum energy $V_{\Lambda}$, in order to yield an extremely small value in the sum $V_{cosm} = V_{vac} + V_{\Lambda}$.
For a historical overview of the quantum vacuum and cosmological constant problem see Ref. \cite{Rugh} and for a pedagogical but technical review see Ref. \cite{Cosmological_constant}.

\section{Electroweak phase transition}
\label{EPT}
\begin{figure}[htb]
\begin{center}
\includegraphics[height=1.7in,width=2.3in]{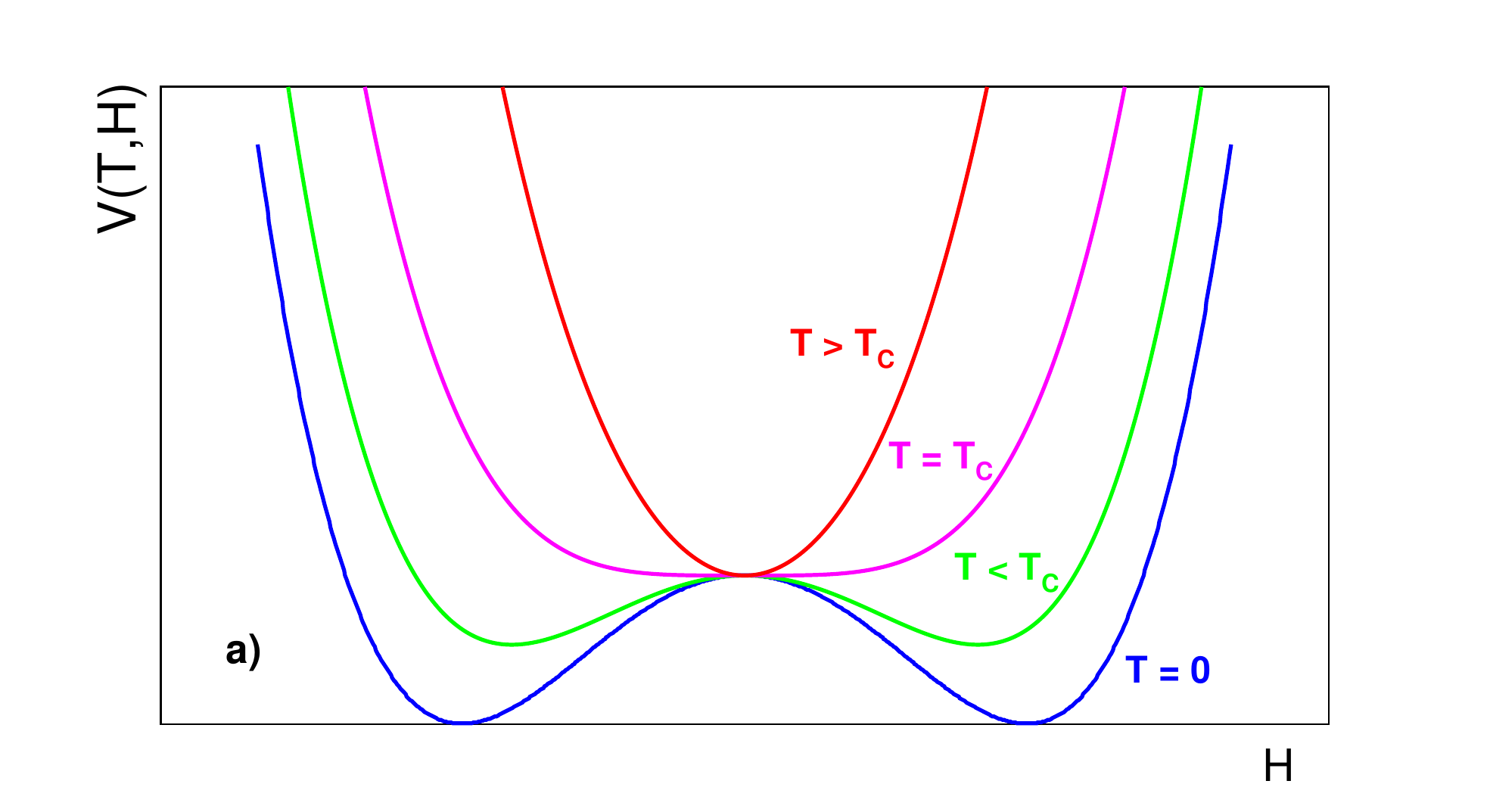} 
\includegraphics[height=1.7in,width=2.3in]{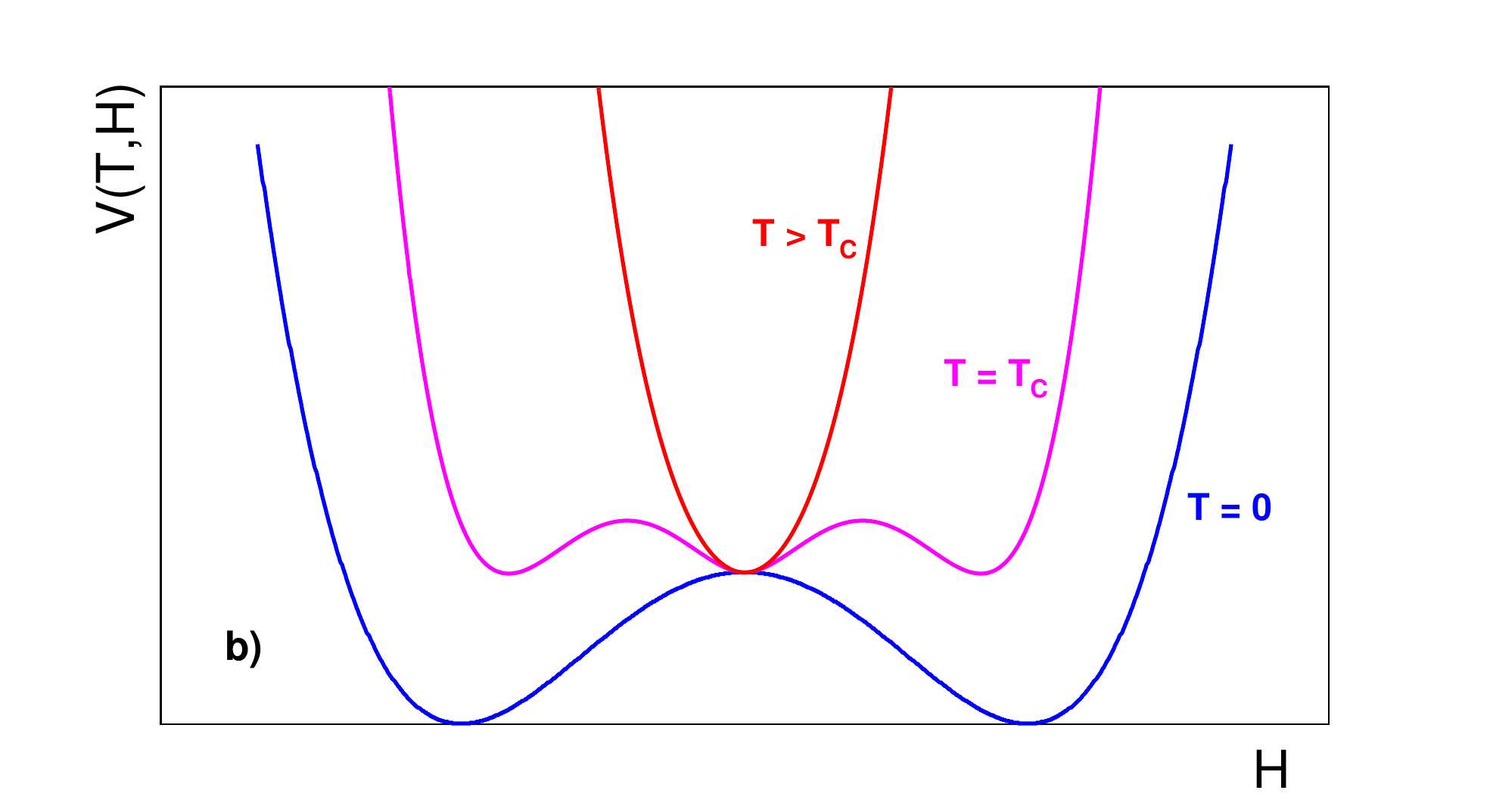}
\caption{Higgs potential for several temperatures compared with the critical temperature $T_c$ for a) second order phase transition. b) first order phase transition.}
\label{Phase_transition}
\end{center}
\end{figure}

Elementary particles became massive in the very early Universe when the Higgs potential took the form shown in Fig.\ref{Higgs_potential_fig}b as a result of the electroweak phase transition which occured about a nanosecond after the Big Bang at the critical temperature $T_c \sim 160$ GeV.
To reconstruct the Higgs potential at that time,  calculations are performed within the so-called finite temperature effective field theory.
The dominant temperature correction to the classical Higgs potential of Eq.\ref{Higgs_potential} is proportional to $T^2$, leading to the effective potential  \cite{Higgs_potential_T2}
\begin{eqnarray}
\label{Higgs_potential_T}
V(T,H) & = & V(H) + b\: T^2 H^2 \;=\; (\mu^2 + b\: T^2) H^2 + \lambda H^4,
\end{eqnarray} 
where $b$ is a coefficient which depends on the couplings of the SM particles to the Higgs field. The combination $\mu^2+b\: T^2 = - \lambda v^2 + b\: T^2$ plays the role of the $\alpha$ parameter of the ferromagnet energy in Eq. \ref{freeenergy}. It is positive for $T > T_c$, zero for $T = T_c$ (yielding $T_c = \sqrt{\lambda v^2/b}$) and negative for $T < T_c$.  
This potential is shown in Fig.\ref{Phase_transition}a.
For $T > T_c$ the potential is symmetric with the minimum at $H = 0$ (the vacuum of the very early Universe).
At $T = T_c$ the valley becomes very flat and as soon as $T < T_c$, the potential develops two minima, one at $H > 0$ and the other one at $H < 0$. Finally, for $T = 0$ the minima (which move away from $H = 0$ during cooling) 
arrive at $H = \pm v/\sqrt{2}$ and the potential
becomes identical with the one in Fig.\ref{Higgs_potential_fig}b. This kind of phase transition is called second order (like the ferromagnetic phase transition).

Another possibility is the first order phase transition (like the boiling of water) depicted in Fig.\ref{Phase_transition}b. In this case we have three degenerate minima at $T = T_c$: the original one at $H = 0$ and the two minima at nonzero $H$ separated by a barrier from the central minimum. At some $T < T_c$ the Universe tunnels through the barrier and takes its position at one of the two minima with nonzero value of the Higgs field.
The first order scenario could be realized through the next order temperature correction to $V(T,H)$ of Eq. \ref{Higgs_potential_T}.
As calculations show \cite{lattice}, the electroweak phase transition is neither first nor second order but a crossover\footnote{The three transitions differ in the temperature dependence of the order parameter: there is a discontinuity for the first order transition but no jump for the other two. The crossover transition, unlike the second order one, is continuous also in the first derivative of the order parameter.} in the SM, however, it could be a first order transition in models beyond the SM.
 

The nature of the phase transition is very interesting from the cosmological point of view.
The first order electroweak phase transition generates gravitational waves, which could
potentially be detected by a space-based gravitational wave interferometer \cite{Higgs_potential_T1}.
The first order transition is also required in order to explain the observed matter-antimatter asymmetry in the Universe.



\section{Naturalness problem}
Difficulties known as the naturalness problem arise when we try to calculate the Higgs potential in quantum field theory (at $T=0$). 
The calculation involves sum of many contributions, such as energy from the quantum fluctuations of the Higgs field itself,
energy from the fluctuations of the top quark field, $W$ and $Z$ fields and so on for all the fields which interact with the Higgs field.
The contributions fall into one of the two classes: i) the SM and ii) physics beyond the SM (BSM physics) such as quantum theory of gravity or new heavy particles.

For SM fields each contribution can be calculated at least approximately for any Higgs field value $H$ between zero and $v_{max}$, and for all quantum fluctuations with energy less than about $\Lambda \sim v_{max}$. The scale $v_{max}$  represents the boundary up to which we can apply the SM \cite{Strassler}. At and above this boundary the BSM physics has to be included. The value of $v_{max}$ is larger than about $10^3$ GeV but we do not know whether it is $10^4$ GeV or $10^{19}$ GeV.

For $H \ll \Lambda$, the SM contribution is dominated by a term proportional to $\Lambda^2\:H^2$
\cite{Coleman_Weinberg,Veltman}  
\begin{eqnarray}
\label{Higgs_potential_top}
V^1  & = & \frac{3}{16 \pi^2 v^2}\big(-4 m_t^2 + \frac{M_h^2}{2} + 2 M_W^2 + M_Z^2\big) \Lambda^2 H^2 \\
     & + & O(H^4\ln{\frac{H}{\Lambda}}) + O(H^4) + ...     \nonumber
\end{eqnarray} 
where $m_t, M_W, M_Z$ are the top quark, W and Z boson masses. Due to its large mass, the top quark contributes the most.

The contribution of BSM physics is not known but for energies well below the scale $\Lambda \sim v_{max}$ it can be absorbed into parameters $\mu_0^2, \lambda_0$
of the $V^0$ potential \cite{Bar_Shalom} given by
\begin{eqnarray}
\label{Higgs_potential_tree_level}
V^0 & = & \mu_0^2 \; H^2 + \lambda_0 \; H^4.  
\end{eqnarray}
The so-called bare potential $V^0$ has the same form as the classical potential of Eq.\ref{Higgs_potential} but the bare parameters $\mu_0^2, \lambda_0$ are unknown. 

The theoretical quantum potential, in our approximation $V^Q(H) = V^0 + V^1$, should be equal to the classical potential $V(H)$, at least
in the experimentally probed region around $H_{vac} = v/\sqrt{2}$ (Fig.\ref{Higgs_potential_fig}b), 
\begin{eqnarray}
\label{sum_V0_V1}
V(H) & = & V^Q(H) = V^0 + V^1.
\end{eqnarray}
Note, however, that the quantum potential differs from the classical one for $H \gg H_{vac}$ (see the next section).
\begin{figure}[htb]
\begin{center}
\includegraphics[height=2.2in]{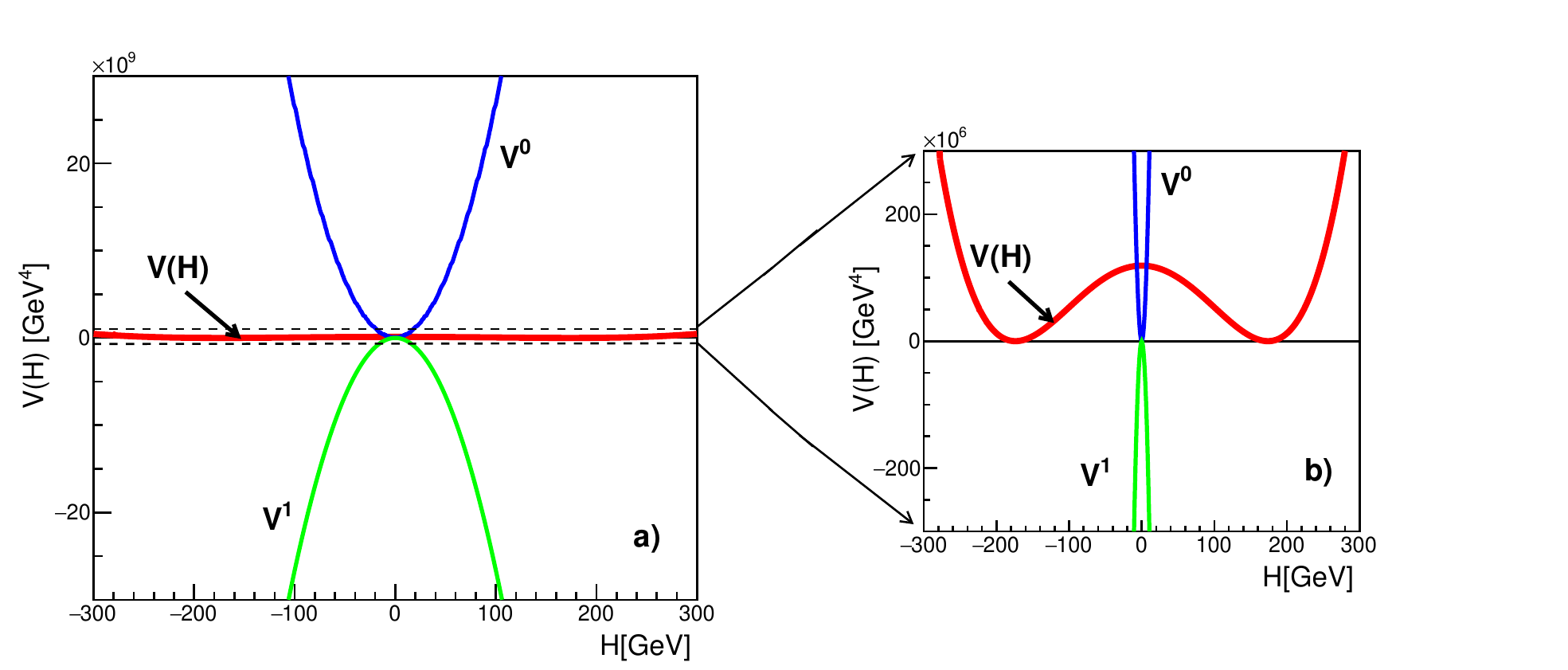}
\caption{a) The tiny classical Higgs potential $V(H)$ (red) appears almost flat compared to the huge contribution $V^1$ from the SM (green). The new physics contribution $V^0$ (blue), 
although apparently unrelated to $V^1$, should equal $-V^1$ with high precision. b) Zoomed in version of a). $\Lambda = 10^4$ GeV.}
\label{Naturalness_Melo_fig}
\end{center}
\end{figure}

The naturalness problem appears if $\Lambda \gg 10^3$~GeV.
The terms of Eq.\ref{Higgs_potential_top} then make $|V^1| \gg V(H)$, see Fig.\ref{Naturalness_Melo_fig}.
In turn, $V^0$ also has to be very large and almost equal in size to $V^1$ (but with opposite sign) in order to secure cancellation and 
yield small $V(H)$. If $\Lambda$ was around the Planck scale $\sim 10^{19}$~GeV, the difference between $V^0$ and $|V^1|$ would have to be only one part
in $10^{34}$ or so. This is very unnatural (fine-tuning) since $V^0$, which is due to BSM physics, appears in no way related to $V^1$, which is due to the SM.
Why should a quantum theory of gravity plus anything else new give a contribution equal to the SM contribution up to 34 decimal places?
The probability of this happenning accidentally is $10^{-34}$. 

We stress that all this is true only to the degree that $\Lambda \sim v_{max}$ is much larger than $10^3$ GeV. For $v_{max}$ close to $10^3$ GeV the fine-tuning is less significant.

The naturalness problem has been a driving force in particle physics for decades. One class of solutions (such as supersymmetry or Little Higgs theories) concentrates on symmetries which predict new particles and ensure that individual contributions to $V^1$ are related - they come in pairs (SM particle and its new partner) of almost equal size and opposite signs leading to $V^1$ which is not much larger than $V(H)$ but rather comparable in size. Another class of solutions (such as technicolor) argues that $v_{max}$ is in fact close to $10^3$ GeV.

In conclusion of this section we recommend a beautiful qualitative discussion of the naturalness problem by M. Strassler \cite{Strassler}.
 
\section{Is our vacuum stable?}
In this section we assume no new physics until the Planck scale $v_{max} \sim 10^{19}$~GeV where quantum gravity becomes important. 

\begin{figure}[htb]
\begin{center}
\includegraphics[height=2.in,width=4in]{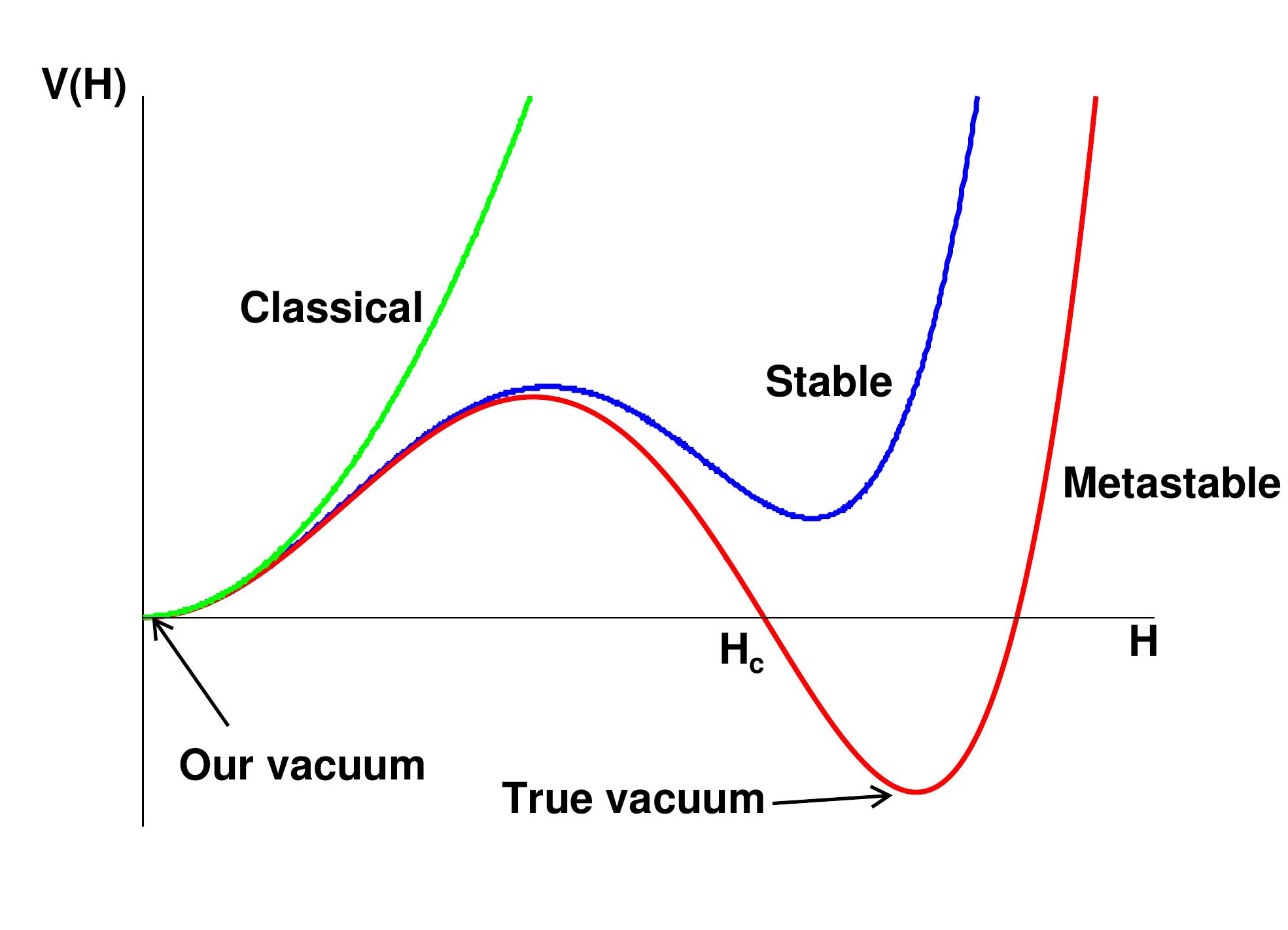} 
\caption{The classical potential (green) and two scenarios for the quantum potential, stable (blue) and metastable (red). In the stable scenario our vacuum is in the global minimum of the potential. 
In the metastable case our vacuum sits in the local minimum while a true, deeper minimum exists. Note: graph not to scale. The local maximum of the classical potential at $H = 0$ is too small to be seen here.}
\label{False_vacuum123}
\end{center}
\end{figure}

The quantum potential $V^Q(H)$ is equal to the classical potential $V(H)$
in the region $H \sim H_{vac} = v/\sqrt{2}$ ({\it cf.} Eq.\ref{sum_V0_V1}).
The full calculation\footnote{This involves many subtle points including the so-called renormalization which fixes the unknown bare parameters $\mu_0^2$ and $\lambda_0$.} 
shows, however, that there is a difference between the classical and quantum potentials for $H \gg H_{vac}$. 
In particular the top quark induces the crucial change in the behaviour of $V^Q(H)$ which starts to fall and might (or might not) become negative at some $H_c$ 
(see the red line in Fig.\ref{False_vacuum123}). We do not know for sure since
the exact answer is very sensitive to uncertainties in the value of the top quark mass $m_t$ and the Higgs mass $M_h$. 

Three scenarios are possible: stable, metastable and unstable. In the stable scenario (blue line in Fig.\ref{False_vacuum123}) the Universe lies safely in our current vacuum which is the global minimum of the potential at $H_{vac} = v/\sqrt{2}$. In the metastable case  (red line) there is another, true minimum, and the Universe might tunnel out from our local minimum (false vacuum) into the true vacuum state with a small probability.
The unstable scenario looks qualitatively like the metastable one except that there is a significant probability for the Universe to tunnel to the true vacuum within its own age.
The most recent SM calculations \cite{Giudice,Branchina} indicate that the metastable scenario might apply - there is a value of the Higgs field, $H_c \sim 10^{11}$ GeV, beyond which the Higgs potential becomes negative and the second, true global minimum develops at large Higgs field values, $H \sim 10^{17}$ GeV \cite{Kobakhidze2}. The height of the barrier between the two minima is as high as $\sim 10^{39}$ GeV$^4$. Under the stated assumption of no new physics until the Planck scale, our Universe appears to be ready to tunnel to the true minimum with catastrophic consequences\footnote{For one thing, the masses of particles would drastically change with the new $H_{vac}$.}, albeit with very low probability.

As noted, the border between stability, metastability and instability is extremely sensitive to  $m_t$ and $M_h$. While our vacuum appears to sit in the region of metastability, it is just within $2 \sigma$ from the stability region \cite{Giudice} and the conclusion is not definitive yet. 
For a calculation of the tunnelling probability of our Universe to the true vacuum, see for example, Ref. \cite{Isidori}.

The decay of our metastable vacuum could be, at least in principle, catalysed by cosmic ray collisions which could lead to an increased tunnelling probability. This question was studied
by Ref. \cite{Enquist}. Their results also indicate that "vacuum decay is very unlikely to be catalysed by particle collisions in accelerators; the total luminosities involved are simply far too low".

\section{Conclusions}
The Higgs field with its nonzero average value and large quantum contributions to its potential is unique among the quantum fields of the SM. The field is responsible for masses of elementary particles, it seems to contribute a huge amount of energy to the vacuum, it may have played (together with some physics extending SM) a significant role in generating the matter-antimatter asymmetry in the Universe. 

The naturalness problem is one of the most important open questions in physics and the possibility of our vacuum being a false vacuum could trigger the phase transition of the Universe to the true minimum of the Higgs potential. All these phenomena rest on a fundamental property - the Higgs field is a scalar (spin $0$) field, the only one in the SM. With other scalar fields contemplated by cosmologists,
the inflaton field and possibly the dark energy field, we may just be entering a new Scalar era.

\section*{Acknowledgement}
I would like to thank the members of the International Particle Physics Outreach Group (IPPOG) for discussions which inspired me to write this paper.


\end{document}